\documentclass[%
 reprint,
 amsmath,amssymb,
prb,
]{revtex4-2}
\usepackage{dcolumn}
\usepackage{balance}
\usepackage{gensymb, graphicx, physics}
\usepackage[usenames, dvipsnames]{xcolor}

\begin{document}
\title{A Gaussian Approximation Potential for amorphous Si:H}

\author{Davis Unruh$^1$, Reza Vatan Meidanshahi$^2$, Stephen M. Goodnick$^2$, G\'abor Cs\'anyi$^3$, Gergely T. Zim\'anyi$^1$}
\affiliation{$^1$Physics Department, University of California, Davis, Davis, CA 95616}
\affiliation{$^2$School of Electrical, Computer and Energy Engineering, Arizona State University, Tempe, AZ 85287-5706}
\affiliation{$^3$Engineering Laboratory, University of Cambridge, Trumpington Street, Cambridge, CB2 1PZ, United Kingdom}

\date{\today}

\begin{abstract}

Hydrogenation of amorphous silicon (a-Si:H) is critical for reducing defect densities, passivating mid-gap states and surfaces, and improving photoconductivity in silicon-based electro-optical devices.
Modelling the atomic scale structure of this material is critical to understanding these processes, which in turn is needed to describe c-Si/a-Si:H heterjunctions that are at the heart of the modern solar cells with world record efficiency. 
Density functional theory (DFT) studies achieve the required high accuracy but are limited to moderate system sizes a hundred atoms or so by their high computational cost.
Simulations of amorphous materials in particular have been hindered by this high cost because large structural models are required to capture the medium range order that is characteristic of such materials. 
Empirical potential models are much faster, but their accuracy is not sufficient to correctly describe the frustrated local structure.
Data driven, ``machine learned'' interatomic potentials have broken this impasse, and have been highly successful in describing a variety of amorphous materials in their elemental phase.
Here we extend the Gaussian approximation potential (GAP) for silicon by incorporating the interaction with hydrogen, thereby significantly improving the degree of realism with which amorphous silicon can be modelled.
%
%
We show that our Si:H GAP enables the simulation of hydrogenated silicon with an accuracy very close to DFT, but with computational expense and run times reduced by several orders  of magnitude for large structures.
We demonstrate the capabilities of the Si:H GAP by creating models of hydrogenated liquid and amorphous silicon, and showing that their energies, forces and stresses are in excellent agreement with DFT results, and their structure as captured by bond and angle distributions, with both DFT and experiments.
%
%
\end{abstract}

\maketitle

\section{Introduction}

Hydrogenated amorphous silicon (a-Si:H) is a widely used material, with applications ranging from thin-film transistors \cite{thompson1987} to solar cells \cite{wronski1976, Bertoni2019}. Created using plasma-enhanced chemical vapor deposition of silane gas (SiH$_4$) \cite{chittick1969}, a-Si:H has far fewer defects than pure amorphous silicon due to the passivating role of the H atoms, the concentration of which typically ranges from 6 to 18 at. \%. As a result  a-Si:H has demonstrably superior electrical transport properties and photoconductivity \cite{leComber1970, spear1974} in comparison to pure amorphous silicon. Therefore the role of H in a-Si:H films is crucial for obtaining effective devices. 

Remarkably, several aspects of a-Si:H remain the focus of research:  understanding the local atomic interactions, the classes of defects and their formation and statistics, and the long-term structural dynamics all pose questions that are unsettled. Of particular interest is the light-induced defect formation, known as the Staebler-Wronski effect \cite{staeblerwronski}, that is directly relevant for photovoltaic applications. Another important issue is the degradation of hydrogen passivation at the amorphous/crystalline interface in silicon heterojunctions (Si HJs) \cite{Bertoni2019, Holovsky2020, soldeg}, which is probably a key driver of the enhanced degradation \cite{NRELdegradation} of the efficiency world record holder Si HJ solar cells\cite{HJWorldrecord}. This enhanced degradation is one of the main factors that slows down the market acceptance of these extremely promising Si HJ solar cells. 

All of the above challenges demonstrate the need for theoretical and simulation-based research efforts to provide crucial guidance and insight. Atomistic simulation methods are the primary tools for materials modeling. The most accurate of these methods are those based on ``first princples'', in this case,  electronic-structure theory. The most prominent method is density functional theory (DFT)\cite{conceptual_DFT}, which is highly accurate, but at the cost of significant computational expense and unfavorable scaling behavior (typically cubic, although linear scaling versions exist with large prefactors\cite{onetep_review}). With rare and heroic exceptions, DFT techniques are typically used to simulate systems with no more than a few hundred atoms. Larger scale simulations of many hundreds/few thousand atoms, or disordered systems that require simulating many thousands of samples, are typically undertaken using parameterized empirical interatomic potentials within Molecular Dynamics (MD) methods\cite{mikefinnis2003}. These potentials typically have simple functional forms, and are usually optimized to reproduce a few key material characteristics on a limited set of structures with good accuracy. These functional forms are parametrized to strike a balance between obtaining good accuracy for the selected observables for the selected structures, and transferability to other observables and/or other structures for which the potential was not fitted to. The limited number of parameters of these models make it very hard to satisfy both of these goals. 

Several interatomic potentials have been developed for MD simulations of pure silicon phases and achieved reasonable accuracy \cite{Si-MD-Review-1992,Cook-1993,Goddard-2003}. However, as discussed further below, even for Si there was clearly room for improving the correspondence with experimental and DFT results. Remarkably, there have been notably fewer papers on Si-H interatomic potentials, even though these are essential for accurate modeling of solar cells. Ref.\cite{Soukoulis-1992} reported limited correlation between a-Si:H MD simulations using the Stillinger-Weber interatomic potential for the Si-H pair-correlation function with experiments. Ref.\cite{Ohira-1996} reported moderate correlation between a-Si:H MD simulations of the Si-H pair correlation function using the Tersoff potential with experiments. 
Ref.\cite{Tuttle-1996} reported discrepancies between a-Si:H MD simulations using various MD potentials and experiments regarding the radial distribution function and vibrational frequencies. Ref. \cite{Murty-Atwater-1995} reported that the hydrogen pairing tendencies observed in Ref. \cite{Boland-1993} were not adequately captured by MD simulation. Finally, Ref.\cite{Sakiyama-2006} reported that MD simulations of reaction processes were inadequate with existing MD potentials, and went on to suggest improvements. 

In the last decade the adoption of machine-learning (ML) methods \cite{GAP1, GAP2, PRX, Thompson2015, Chen2017, Behler2007, Behler2011, Behler2016, Kocer2019, ShapeevMTP2016, KresseVASPGAP, Tkatchenko2017, ANI, ACE} has changed the above picture, particularly for the simulation of hard materials. These ML-based methods construct highly flexible non-parametric interatomic potentials by training them on a DFT-computed energies and gradients of a wide variety of structures, and can deliver near DFT level accuracy for {\em similar} structures at a computational cost that is thousands of times less than that of DFT for moderate sizes systems. Since interatomic potentials scale linearly with system size, the savings can reach factors of millions for large structures or when using expensive hybrid correlation functionals. For recent reviews of these methods and their application to materials simulation, see Refs. \citenum{Csanyi_AdvMater,  deringer_chemrev}. 


The advantage of ML models in general, and ML-based potentials in particular, is the flexibility of their functional form, which enables them to accurately interpolate the potential energy of structures in a broad training database, covering a wide variety of structures and phases. However, the nature of this flexibility also leads to the ML models to be increasingly less accurate for structures that are  farther away from the training data set. Thus the disadvantage of ML-based potentials is that their limited transferability or extrapolation. The key to success is to transform the problem of fitting the total energy of large systems into smaller subtasks for which interpolation suffices: for extended materials this is achieved by the ansatz that the total energy is a sum of local terms, {\em site or atomic energies}, which are only functions of the local environment. Creating a truly ``general-purpose'' potential thus requires a very large training database that well covers all relevant {\em local environments}. When the potential is used, it will be accurate as long as the configurations encountered consist of local environments that are near those in the database. If radically new local environments are encountered (e.g. because of phase unanticipated transitions or the system is taken to a pressure or temperature range far outside from where the training structures came from), accuracy will be severely compromised. Note that the notion of ``nearness'' used above in practice is defined by the architecture of the ML model and the types of regularisation that are used in obtaining the optimal parameters. 

In this work we use the Gaussian Approximation Potential (GAP) framework \cite{GAP1,GAP2}. GAP is based on the Gaussian process regression (also known as kernel regression) methodology, typically using the Smooth Overlap of Atomic Positions (SOAP) kernel to describe the similarity between local environments in materials.\cite{GAP2} One advantage of kernel regression is that the flexibility of the functional form naturally grows with the amount of training data. The general approach, which has recently been reviewed in detail\cite{deringer_chemrev}, is to fit a potential to an initial training set of total energies, forces and stresses calculated using a given electronic structure method (typically density functional theory), and then use this potential to explore a wide range of structures using a variety of algorithms including molecular dynamics, geometry optimisation, etc. The training data set is then iteratively broadened, particularly with configurations where the fit is not yet good enough, until the desired accuracy is reached for the configurations and observables of interest.  The ultimate validation of the potential is by testing macroscopic observables using samples generated by the potential and not explicitly used in the training.

A GAP model  has already been developed to describe pure Si\cite{PRX}, including many crystalline and amorphous phases and has been used to study the structure of a-Si\cite{deringer_aSi_2018}, and its phase transitions under pressure in unprecedented detail and precision\cite{aSiNature}. However,  a-Si/c-Si heterojunction (HJ) solar cells contain significant amount of hydrogen, most often in the 10-15\% range. It is well-known that existing interatomic potentials of Si:H are unable to capture experimentally measurable quantities with sufficient precision, as discussed earlier \cite{Soukoulis-1992,Ohira-1996,Tuttle-1996, Murty-Atwater-1995,Boland-1993,Sakiyama-2006}, including partial radial distribution functions, pairing tendencies and reaction processes. Therefore, the effort to describe, analyze and then mitigate the problematic excess degradation of these HJ cells requires the development of a GAP model for Si:H, which we report on in this paper.  We validate our Si:H GAP on hydrogenated liquid Si and amorphous Si, which are the two most relevant phases since a-Si:H is typically modelled using melt-quench simulations of liquid Si:H. To gauge the utility of our Si:H GAP, we show how it qualitatively outperforms state-of-the-art parameterized interatomic potentials by delivering excellent agreement with DFT calculations and experimental results. This Si:H GAP will be key for future research by qualitatively enhancing the precision of MD simulations of hydrogenated silicon solids to DFT levels, enabling a wide variety of electro-optical simulations, including simulations of a-Si:H/c-Si heterojunction solar cells.

\section{The GAP platform}

In this project we train a Gaussian approximation potential (GAP) on a set of structures and accompanying energies, forces and stresses, all compute using DFT. The GAP is constructed as the sum of a purely repulsive ``core'' potential and a Smooth Overlap of Atomic Positions, SOAP kernel. The repulsive potential is built with cubic splines that are fitted to the interaction of pairs of atoms in vacuum as computed by DFT. These pair potentials are built for all three relevant atomic pairs: Si-Si, Si-H, and H-H. Including these repulsive core pair potentials serves a dual purpose. First, a large fraction of the interaction energy between atoms can in fact be described by a pair potential, which describes exchange repulsion at close atomic distances, and some effects of chemical bonding at far distances. Second, the repulsive portion of the potential is difficult to capture using the same kernel function that is appropriate to describe valence bonding due to the steepness of the energy curve at close approach, and thus capturing it with a pair potential with a piece-wise spline enhances overall numerical efficiency significantly.

The total GAP energy for the system is generated as a sum of the repulsive pair  potential and the many-body SOAP kernel which is constructed as a linear sum over kernel basis functions:
\begin{equation}
    E = \sum_{i<j}{V^{(2)}(r_{ij})} + \sum_i\sum_{s}^{M}{\alpha_s K(R_i, R_s)}
\end{equation}
where $i$ and $j$ index  the atoms in the system, $V^{(2)}$ is the 2-body repulsive pair potential, $r_{ij}$ is the distance between atoms $i$ and $j$, $K$ is the SOAP kernel basis function, and $R_i$ is the collection of relative position vectors corresponding to the neighbors of atom $i$ which is termed a \textit{neighborhood}. The $s$ sum ranges over the set of $M$ \textit{representative} atoms, selected from the input data set, whose neighborhoods are chosen to serve as a basis in which the potential is expanded. The coefficients $\alpha_s$ are determined by a regularised linear regression of the energies, forces and stresses of the system computed with Eq.(1) that are parametrized by these $\alpha_s$s values, and compared to the corresponding energies, forces and stresses computed by DFT for the same structures. This comparison results in a set of linear equations that are solved for the coefficients $\alpha_s$ using Tikhonov regularisation. The representative neighborhoods over which Eq. (1) is evaluated are selected by choosing basis neighborhoods which are maximally dissimilar to each other (using CUR matrix decomposition of the matrix corresponding to all the neighbourhoods in the training data set), such that the variety of the entire set of possible neighborhoods can be well represented by interpolation over a small number of these basis neighborhoods. The details of all these algorithms and their general application to fitting a GAP model have recently been reviewed in Ref. \citenum{deringer_chemrev}.

The SOAP kernel $K(R_i, R_s)$ characterizes the similarity between two neighborhoods: it is maximal when the two neighborhoods are identical, and it is smaller and smaller for more and more different neighborhoods. To quantify the similarity of two neighborhoods, each neighborhood $R_i$ of atom $i$ is represented by a set of corresponding neighbor densities, a separate one built for each kind of element in the neighbourhood:

\begin{equation}
    \rho^{z}_i(\mathbf{r}) = \sum_{i'} \delta_{zz_{i'}}{f_\mathrm{cut}(r_{ii'})e^{-(\mathbf{r} - \mathbf{r}_{ii'})/2\sigma^2_\mathrm{atom}}},
\end{equation}
where $z$ correspond to the element whose density around atom $i$ is constructed and  the $\delta_{zz_{i'}}$ factor selects neighbours corresponding to element $z$ to be included in the sum. In the present case, there are two elements (Si and H), so two neighbour densities are built around each atom. The calculation of the SOAP kernel starts by integrating the overlap of the two neighbor densities:
\begin{equation}
    \tilde{K}(R_i, R_j) = \sum_{zz'} \int_{\hat{R}\in \mathrm{SO(3)}} d\hat{R}\left| \int d\mathbf{r} \rho^{z}_i(\mathbf{r})\rho^{z'}_j(\hat{R}\mathbf{r}) \right|^2
\label{eq:kernel_integral}
\end{equation}
Next, the integral $\tilde{K}$ is normalized by the self-overlaps. Finally, it is raised to an integer power and has a suitable hyperparameter prefactor assigned:
\begin{equation}
    K(R_i, R_j) = \delta^2\left| \frac{\tilde{K}(R_i,R_j)}{\sqrt{\tilde{K}(R_i, R_i)\tilde{K}(R_j, R_j)}} \right|^\zeta
\end{equation}
In practice, the neighbor densities are represented by an expansion in a basis of spherical harmonics and radial functions, and the integral in \eqref{eq:kernel_integral} is efficiently evaluated as a scalar product. A detailed review of SOAP and similar neighbour density based representations, descriptors and kernels is in Ref.~\citenum{Csanyi_ChemRev}. The particular application to pure Si was comprehensively described in Ref. \citenum{PRX}, which we refer to for further details and validation.

\section{Parametrising the Si:H GAP}
\subsection{Outline of training and reference structures}

ML models in general, and GAP models in particular are trained using two sets of structures. First, GAP is trained by fitting to DFT-calculated energies, forces and stresses on the structures of an adaptively created training set. This training set is  iteratively expanded by creating new structures with simple MD simulations (such as a structural relaxation or a brief thermal protocol) that use the current version of the GAP model, starting from structures already contained within the training set. We then carry out DFT and GAP tests on the new structures, and determine the differences between them. Structures for which the comparative fit is poor are added to the next generation training set and the Si-H GAP model is retrained. 

Second, the results of the training procedure are regularly evaluated using a reference set. The GAP is not trained on this ``validation set'' of structures. Instead, the energies, forces, and stresses are calculated with both the GAP and with DFT on the structures within the validation set, and the difference is used to quantify and monitor the progress of the training.  

The structures of the training set and the validation set were assembled by adding H to pure Si structures of representative Si phases and structures. The initial training set included about 150 structures. The validation set included about 110 structures. These pure Si structures were taken from our previous work on a-Si/c-Si interface degradation \cite{soldeg}, from the reference database on which the published Si-only GAP was trained \cite{PRX}, or were generated using the Atomic Simulation Environment \cite{ASE}. The representative phases of Si were: 1) Amorphous Silicon; 2) Liquid Silicon; 3) Diamond Silicon with a vacancy; 4) Diamond Silicon with a divacancy; 5) Diamond Silicon with an interstitial Si; 6) Amorphous/Crystalline Silicon interface structures; and 7) Diamond surface structures, with (100) and (111) orientation. The atomic concentration of added H added was in the 6 - 12 at.\% range for the liquid and amorphous phases and for the interface structures, in the 4 - 8 at. \% range for the bulk diamond phases, and in the 12 - 20 at. \% range for the diamond surfaces. In each of these structure a sufficient amount of H was added to fully passivate all dangling or highly strained bonds. Furthermore, in order to train the Si-H GAP to accurately model hydrogen-related defects, structures with an H atom added or taken away from the fully passivated structures were also added both to the training set and to the validation set. 

Finally, it is noted that the training set also contains an isolated Si atom and an isolated H atom as ``structures''. Isolated here means the structure possesses a large enough unit cell that the atom is effectively in isolation, even though its energy is computed using periodic boundary conditions and the same DFT parameter settings, for consistency. Including these isolated atoms as training structures is essential because the GAP is fit to the binding energy, not the total energy (i.e. the binding energy plus the energy of isolated atoms). 

\subsection{DFT Calculation Details}


The DFT calculations were performed using the Quantum Espresso 6.2.1 software package \cite{QE1, QE2, QE3}, with the key parameters as follows. The Perdew-Burke-Ernzerhof (PBE) exchange-correlation functional was used with periodic boundary conditions \cite{PBE}. The core and valence electron interactions were described by the Norm-Conserving Pseudopotential function. The calculations were performed with Marzari-Vanderbilt electronic smearing \cite{marzarivanderbilt}. This smearing method was chosen as it ensures that the DFT energies and forces are consistent, and the electronic free energy was used as the ``energy'' training target, because its derivaties with respect to the atomic positions are reported as ``forces'' by the DFT code (according to the Hellman-Feynman theorem). An energy cutoff of 42 Ry was employed for the plane-wave basis set, and a Monkhorst-Pack grid method was used to define the k-point mesh which samples the Brillouin-Zone. The k-point spacing was chosen to be $0.2 \mathrm{\AA}^{-1}$.

We also made sure that that the energy/atom values calculated for each structure were accurate within 1 meV/atom. This was tested by verifying that the energy/atom values did not change more than $\pm1$ meV/atom when the cutoff was pushed to exceptionally large values, or the k-point density was substantially increased.

A separate set of parameters was used to perform DFT Born-Oppenheimer molecular dynamics (DFT-BOMD) simulations \cite{car1985unified}. DFT-BOMD was employed to compare the performance of GAP-driven MD simulations versus DFT-driven MD simulations. The DFT-BOMD simulations utilized the PWSCF module of the Quantum Espresso software, using the same PBE exchange-correlation functional as before. 
An energy cutoff of 36 Ry was used for the plane-wave basis set, and the first Brillouin zone was sampled using only the $\Gamma$ point. A Gaussian smearing width of 0.01 Ry was implemented to the density of states to avoid convergence problems with metallic configurations.


\subsection{Fitting the Si-H GAP}

Several hyperparameters were needed to define the SOAP kernels centered on Si and H atoms. These included: 1) $n_\mathrm{max}$ and $l_\mathrm{max}$, the maximum number of radial and angular indices for the spherical harmonic expansion of the neighbor densities; 2) $\delta$, a hyperparameter that set the energy scale of the many-body term in the SOAP kernel; 3) $\zeta$, the exponent used to construct the SOAP kernel; 4) the cutoff radius $r_\mathrm{cut}$ that characterized the radius beyond which the cutoff function within the neighbor densities converges to zero; and its associated transition width $w$ that set the rate of this convergence; 5) $\sigma_\mathrm{atom}$, the smearing parameter for the neighbor density function. Table \ref{tab:first} summarizes the values of these parameters. 

\begin{table}[t]
\begin{center}
 \begin{tabular}{||c | m{0.8cm} | m{0.8cm} | m{0.5cm} | m{0.6cm} | m{0.8cm} | m{0.8cm} | m{0.8cm}||} 
 \hline
 Element & $n_\mathrm{max}$ & $l_\mathrm{max}$ & $\delta$ & $\zeta$ & $r_\mathrm{cut}$ & $w$ & $\sigma_\mathrm{atom}$ \\ [0.5ex] 
 \hline\hline
 Si & 10 & 6 & 3 & 4 & 5.0 & 1.0 & 0.5 \\ 
 \hline
 H & 9 & 6 & 1 & 4 & 3.5 & 0.5 & 0.4 \\ 
 \hline
\end{tabular}
\caption{Table of Si and H SOAP kernel parameter values}\label{tab:first}
\end{center}

\bigskip

\begin{center}
 \begin{tabular}{|c || c |} 
 \hline
 Phase &  $\sigma_\mathrm{energy}$ [eV/atom]\\ [0.5ex] 
 \hline \hline
 Amorphous Silicon &  0.0015\\ [0.5ex] 
 \hline
 Liquid Silicon &  0.003\\ [0.5ex] 
 \hline
 a-Si/c-Si Interface &  0.003 \\ [0.5ex] 
 \hline
 Diamond Si Phases &  0.001\\ [0.5ex] 
 \hline
 Isolated Atom &  0.0001\\ [0.5ex] 
 \hline
\end{tabular}
\caption{Table of $\sigma_\mathrm{energy}$ values for each structure type.}\label{tab:second}
\end{center}
\end{table}

The fitting also required choosing regularization parameters for each kind of target data: $\sigma_\mathrm{energy}$ for energies, $\sigma_\mathrm{force}$ for force components, and  $\sigma_\mathrm{virial}$ for virial stress components. These regularisation parameters can be set at different values for different 
parts of the training set. These parameters represented the accuracy of the GAP to be reached with the fitting procedure. The relative values of these same parameters for different parts of the training set also determined the relative weight each phase/structure is has in the loss function. It is possible to input these values in the potential in a number of different ways, including using the same value for all structures, or as the same value for all structures in a single phase, or ``structure type'', for instance, liquid Si. We ended up using a combination of these methods: we assigned the regularisation values on a structure-by-structure basis. The $\sigma_\mathrm{energy}$ values were chosen to be the same for all structures in a given phase, see Table \ref{tab:second}. In some detail, $\sigma_\mathrm{force}$ was defined on a per-atom basis as:
\begin{equation}
    \sigma_\mathrm{force} = 
    \begin{cases}
        0.1 & ,\abs{F} < 2.0 \,\mathrm{eV}\mathrm{\AA}^{-1} \\
        0.05 \abs{F} & ,\abs{F} \geq 2.0 \,\mathrm{eV}\mathrm{\AA}^{-1}
    \end{cases},
\end{equation}
where $\abs{F}$ was the magnitude of the force vector on that respective atom. $\sigma_\mathrm{virial}$ was defined on a per-structure basis as:
\begin{equation}
    \sigma_\mathrm{virial} = 
    \begin{cases}
        0.025N_\mathrm{atoms}  & ,\max{\left(\abs{\tau_i}\right)} < 1.0 \,\mathrm{eV} \\
        0.025N_\mathrm{atoms}*\max{\left(\abs{\tau_i}\right)} & ,\max{\left(\abs{\tau_i}\right)} \geq 1.0 \,\mathrm{eV}
    \end{cases}
\end{equation}
where $N_\mathrm{atoms}$ was the number of atoms in the structure, and $\max{\left(\abs{\tau_i}\right)}$ was the maximal norm of the virial stress tensor.

\subsection{Adaptive Training of the Si-H GAP}

\begin{table}[b]
\begin{center}
 \begin{tabular}{|c || m{0.8\columnwidth} |}
 \hline
 Iteration &  Structure Type\\ [0.5ex] 
 \hline \hline
 1 &  Optimized structures (all phases)\\ [0.5ex] 
 \hline
 2 &  Optimized structures (all phases)\\ [0.5ex] 
 \hline
 3 &  Low T anneal of a-Si:H\\ [0.5ex] 
 \hline
 4 &  High T anneal of liq-Si:H\\ [0.5ex] 
 \hline
 5 &  High T anneal of liq-Si:H\\ [0.5ex] 
 \hline
  6 &  Med T anneal (1100K) of a-Si:H\\ [0.5ex] 
 \hline
  7 &  Heating a-Si:H from 500K to 800K at $10^{13}$ K/s\\ [0.5ex] 
 \hline
  8 &  Heating a-Si:H from 800K to 1100K at $10^{13}$ K/s\\ [0.5ex] 
 \hline
  9 &  Heating a-Si:H from 1100K to 1400K at $10^{13}$ K/s\\ [0.5ex] 
 \hline
  10 &  Heating a-Si:H from 1100K to 1400K at $10^{13}$ K/s\\ [0.5ex] 
 \hline
  11 &  Heating a-Si:H from 800K to 1400K at $10^{12}$ K/s\\ [0.5ex] 
 \hline
  12 &  Added new a-Si:H structures\\ [0.5ex] 
 \hline
  13 &  Add new a-Si:H structures\\ [0.5ex] 
 \hline
  14 &  Added c-Si/a-Si:H interface structures\\ [0.5ex] 
 \hline
  15 &  Added c-Si/a-Si:H interface structures\\ [0.5ex] 
 \hline
  16 & Added new c-Si divacancy structures\\ [0.5ex]
  \hline
  17 & Added new liq-Si:H structures\\ [0.5ex]
  \hline
  18 & Added new c-Si vacancy structures\\ [0.5ex]
  \hline
  19 & Added new c-Si interstitial structures\\ [0.5ex]
  \hline
  20 & Low T anneal of c-Si/a-Si:H interface structures\\ [0.5ex]
  \hline
  21 & Optimization of c-Si/a-Si:H interface structures\\ [0.5ex]
  \hline
  22 & NPT high T anneal of liq-Si:H structures\\ [0.5ex]
  \hline
  23 & NVT high T anneal of liq-Si:H structures\\ [0.5ex]
  \hline
  24 & Quenching liq-Si:H from 2000K to 1500K at $10^{13}$ K/s\\ [0.5ex]
  \hline
  25 & Annealing quenched liq-Si:H structures at 1500K\\ [0.5ex]
  \hline
  26 & Quenching liq-Si:H from 1500K to 1400K at $10^{12}$ K/s\\ [0.5ex]
  \hline
  27 & Added hydrogen passivated c-Si surface (100) and c-Si surface (111) structures\\ [0.5ex]
  \hline
\end{tabular}
\caption{Structure types added in each iteration.}\label{tab:third}
\end{center}
\end{table}

As described in the Introduction, the conventional wisdom is that a ML model needs to be trained on a broad database of structures, containing as many structures as possible, in order to be usable in a general sense. Such an approach requires training on enormous training sets, and its product can end up over-fitted with an excessive number of parameters. As an efficient alternative, we adopted an adaptive training procedure, wherein first the GAP was trained on the structures contained in the training set/database. Subsequently, the GAP was validated on structures outside the training set with DFT measurements of energies, forces, and virial stresses. The validation structures for which the quality of the fits fell below a threshold were then added to the training set, and the GAP was then re-trained. Such adaptive procedures grow the training database in an iterative manner. Using this procedure, it was not necessary to add thousands of structures indiscriminately to the training set, as the adaptive training procedure on its own preferentially gravitated only towards structures that needed to be included into the training set without any external input, thereby avoiding the problem of over-fitting.

We performed all MD simulations using the LAMMPS software package built with QUIP package support \cite{LAMMPS, LAMMPSLINK, QUIP} that can use GAP models. We evaluated the accuracy of the GAP compared to DFT with the error metric of the weighted root mean square error (RMSE). As the targeted accuracy of the GAP varied from structure to structure, set by their regularization parameters $\sigma$, the natural measure of accuracy is not the absolute RMSE value, but its ratio to the regularization parameter $\sigma$. We captured this by weighting the RMSE by $\sigma^2$ as follows:

\begin{equation*}
    RMSE_\mathrm{weighted} = \left( \frac{\sum^N_{i=1} (x_{i,\mathrm{GAP}} - x_{i,\mathrm{DFT}})^2/\sigma^2_i}{\sum^N_{i=1} 1/\sigma^2_i} \right)^{1/2}
\end{equation*}

Here, $N$ is the number of data points, $x_\mathrm{GAP}$ is the value measured by GAP, and $x_\mathrm{DFT}$ is the value measured by DFT, where $x$ can be an energy, a force or a virial stress.

\begin{figure}[t]
\includegraphics[width=0.95\columnwidth]{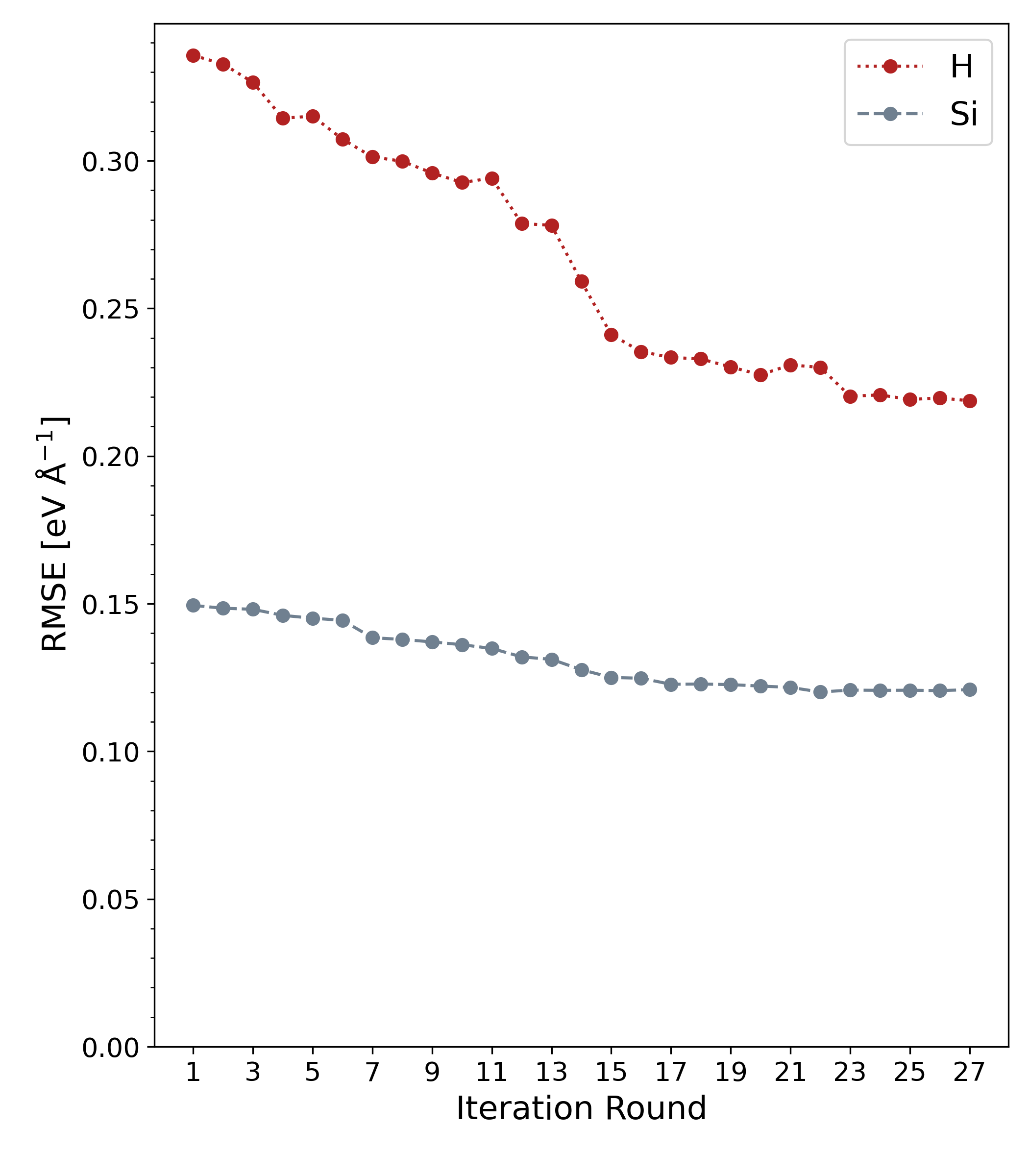}
\caption{Iterative progression of the average of the RMSE of the Si (gray dashed) and H (red dotted) cartesian force components, measured with respect to DFT on the reference structure set.
}
\label{fig:forceComparison}
\end{figure}

We conducted 27 full rounds of iterative training of the Si-H GAP. Fig. \ref{fig:forceComparison} shows the iteration-by-iteration progress of the average of the (unweighted) RMSE of the force components, measured by comparing the Si-H GAP forces to the DFT forces on the reference database, separately for Si and for H. In iterations 1-11 of the above-described adaptive procedure, we added new structures shown in the right column of Table \ref{tab:third}, distinguished by the phase and the preparation method, such as heating or annealing, and doing so across different temperature ranges. Beginning with iteration 12, we started to add new structures to the adaptive training set that were distinguished by their different a-Si:H architecture or structure, such as c-Si/a-Si:H interfaces and different vacancy structures. A few sets of such structures from various phases were added to the training set. Once the new structures were added to the training set, the training of the GAP continued with the newly added structures. Table \ref{tab:third} lists the structures added to the training set over the 27 iterations. The results of this 27 iteration training are discussed next. 

Fig. \ref{fig:forceComparison} shows how the force RMSE on the Si atoms and the H atoms evolved over the 27 training iteration. Broadly speaking, every iteration successfully reduced the force RMSE. The force RMSE improved notably in iteration 7 because we changed the method of assigning the regularization parameters $\sigma_\mathrm{force}$. In the first 6 iterations, the regularization parameters $\sigma_\mathrm{force}$ were assigned the same value for all structures in a given phase. From iteration 7, we switched to assigning the  $\sigma_\mathrm{force}$ values for each structure individually.

In addition, in the same iteration we changed the fitting protocol for the regularization parameters $\sigma_\mathrm{force}$ and $\sigma_\mathrm{virial}$ to depend on the magnitude of the atomic force $\abs{F}$, or stress, measured by DFT, as shown in Eq.(5). We changed the regularization because it was natural to relax the fitting for atoms, structures, and phases where the forces and stresses were larger. These two changes reduced the force RMSE and thus increased the accuracy of the GAP considerably, as visible in Fig. \ref{fig:forceComparison}.

To form an overall picture about the efficiency of the training, we recall that the fitting of each quantity $x$ was guided by the corresponding regularization parameter $\sigma_\mathrm{x}$. As captured by the concept of the weighted RMSE, the GAP-simulated quantities were fitted to the DFT-simulated quantities only with a $\sigma_\mathrm{x}$ accuracy, or tolerance. As such, the $\sigma_\mathrm{x}$ values represent the lowest error thresholds typically attainable with this regularisation procedure. We recall that the regularization parameter $\sigma_\mathrm{force}$ was set to $\sigma_\mathrm{force,min} = 0.100 \,\mathrm{eV}\mathrm{\AA}^{-1}$ for weaker forces $F < F_{th}=2.0 \,\mathrm{eV}\mathrm{\AA}^{-1}$, and proportionally higher for stronger forces $F > F_{th}$. Visibly, the training successfully reduced the (unweighted) RMSE for the forces on the Si atoms from 50\% above $\sigma_\mathrm{force,min}$ to only 20\% above $\sigma_\mathrm{force,min}$. Moreover, since the forces $F$ often exceeded the threshold value $F_{th}$, the effective $\sigma_\mathrm{force,eff}$ was in fact higher than $\sigma_\mathrm{force,min}$. Therefore, the training brought the accuracy of the RMSE even closer than 20\% to the effective $\sigma_\mathrm{force,eff}$.  Thus, the training of the Si-H GAP substantially improved the precision of the calculated force values, and made the Si-H GAP very close to reaching the theoretical limit of its precision. 

The same training also reduced the RMSE for the forces on the H atoms from about 0.34 $\mathrm{eV}\mathrm{\AA}^{-1}$ down to 0.22 $\mathrm{eV}\mathrm{\AA}^{-1}$. Just like for the Si atoms, we place this training into context. For the H atoms, the forces $F$ exceeded $F_{th}$ much more often than for Si atoms. The reason for the higher force RMSE values for H are likely due to locality, as explained and evidenced below.  

We recall that a ``locality limit'' applies to the Si-H GAP because it does not include long-range interactions beyond 5 $\mathrm{\AA}$ for Si and 3.5 $\mathrm{\AA}$ for H. For a discussion of this aspect, see Ref. \citenum{C_GAP}. This locality limit provides a natural procedure to determine an accuracy threshold the Si-H GAP forces can possibly achieve.\cite{C_GAP, GAP1}. We can quantify this limit by a ``locality test'' procedure: a central atom is selected in a simulation cell, and a sphere is defined around it with a radius $r_\mathrm{fix}$. The atoms inside this sphere are kept fixed. The atoms outside this sphere are assigned a velocity corresponding to a high temperature of, e.g., T=2,000K, and evolved with MD using the Si-H GAP for 1 ps. Representative snapshot structures are gathered over the course of this evolution. The force acting on the central atom in these snapshot structures is calculated using DFT. The standard deviation of the distribution of the forces on the central atom over all snapshots is a natural measure of a threshold accuracy attainable by any interatomic potential with the given finite cutoff, since it captures how much the DFT-calculted forces can vary due to configuration changes outside the cutoff. Note that this test is entirely independent of the interatomic potential and the locality that it measures is a quantum mechanical property of the system. 

We performed this locality test on the H atoms in the most disordered phases: the amorphous and the liquid structures. We found that for the appropriate cutoff radius of $r_\mathrm{fix}$ = 3.5 $\mathrm{\AA}$, the above-defined force standard deviation was 0.20 $\mathrm{eV}\mathrm{\AA}^{-1}$. Adopting this result of the locality test as the attainable value of the force RMSE, we conclude that the 27-round training of the Si-H GAP reduced the force RMSE of the H atoms from 70\% above the locality limit to only 10\% above. 

\begin{figure}[t]
\includegraphics[width=\columnwidth]{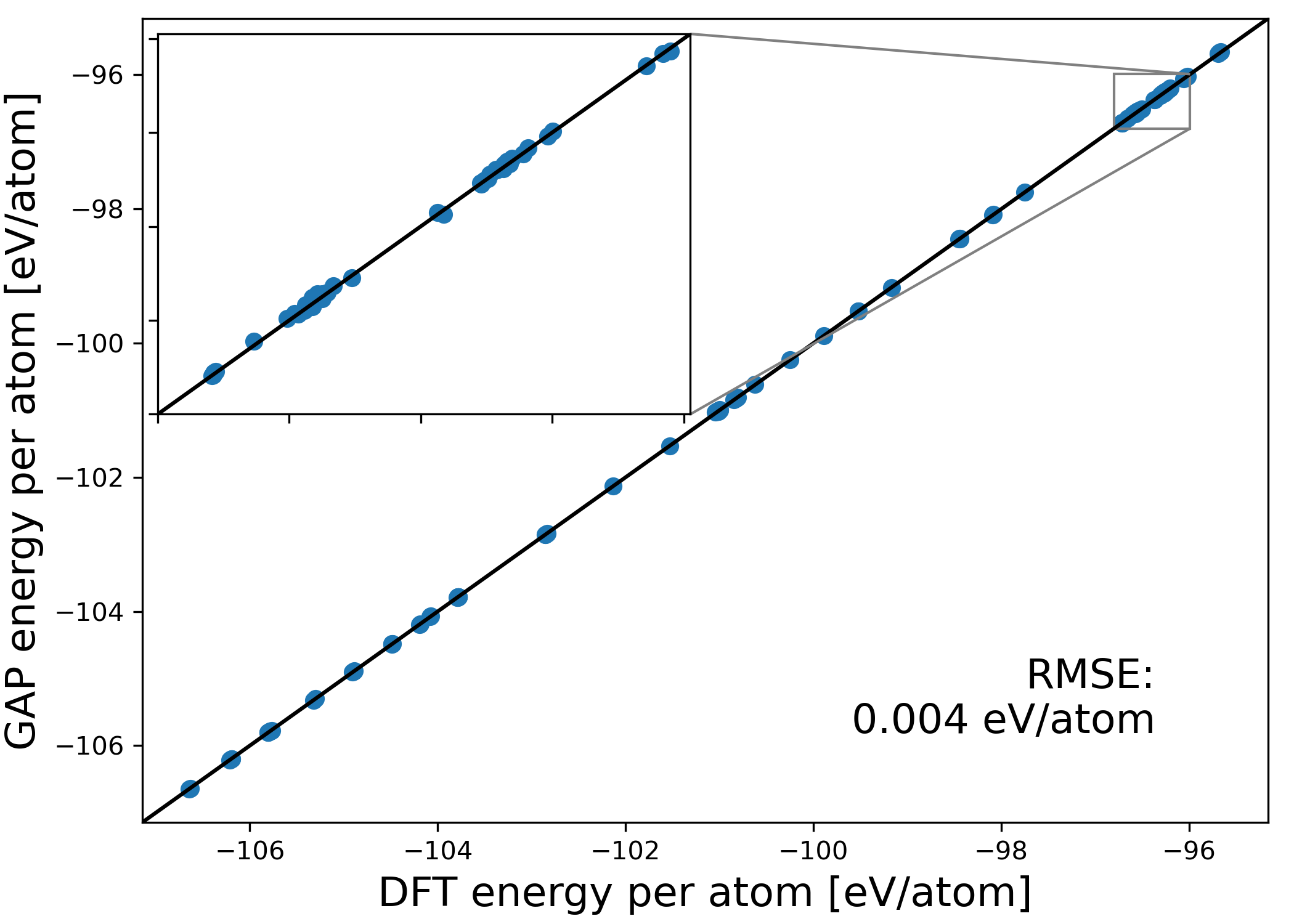}
\caption{Correlation between GAP and DFT measurements of the energies/atom on the reference set of structures, after iteration 27.}
\label{fig:energyComparison}
\end{figure}

Next, Fig. \ref{fig:energyComparison} shows the correlation between the energies/atom computed by the Si-H GAP and by DFT, as measured on the reference set after 27 iteration. Our adaptive method reached an RMSE of this correlation curve to be as little as 4 meV/atom, close to the precision of out DFT  calculations, 1 meV/atom.  

\subsection{Analysis and validation of the Si-H GAP}

We will now analyze the usability of our GAP model by conducting MD simulations with it, calculate experimentally observable quantities, and comparing the results to those obtained by using DFT and by experiments. While our broad goal was to create a universally useful Si-H GAP, our primary motivation and intended use for this Si-H GAP was to create highly accurate a-Si:H structures via a melt-quench procedure from the liquid Si:H, and to subsequently use it to measure defect creation/annihilation energy barriers via the nudged elastic band method. Therefore, we will focus our analysis of the Si-H GAP on the liquid phase liq-Si:H; on the amorphous phase a-Si:H; and on some representative defect structures. 

There are several different metrics for evaluating the``realness'' of the resulting structures that can be experimentally observed and tested. These include: 1) the radial distribution function (RDF) for mono-atomic systems; or the partial pair correlation function for multi-atomic systems; 2) the bond length distributions and the bond angle distribution; 3) the excess energy/atom, measured relative to c-Si; and 4) the vibrational spectra. The relative importance of these experimental metrics is the subject of ongoing debate, with different proponents arguing in favor of either the RDF, or the vibrational spectra \cite{metrics}. It is significantly easier to measure the RDF computationally, but its experimental determination is more difficult, as it requires x-ray diffraction measurements of the structure factor $S(Q)$ out to at least 40$\textrm{\AA}^{-1}$ \cite{Laaziri}. Conversely, it is much easier to measure the vibrational spectra experimentally using Raman and FTIR spectroscopy \cite{ftir}, but it is a substantial challenge computationally. In this paper we report the computation and analysis of the following observables: the partial pair correlation functions, bond angle distributions, coordination statistics, and some defect characteristics, while we leave the study of the vibrational spectra to future work.

\subsubsection{Liquid phase}

\begin{figure}[b]
\includegraphics[width=0.95\columnwidth]{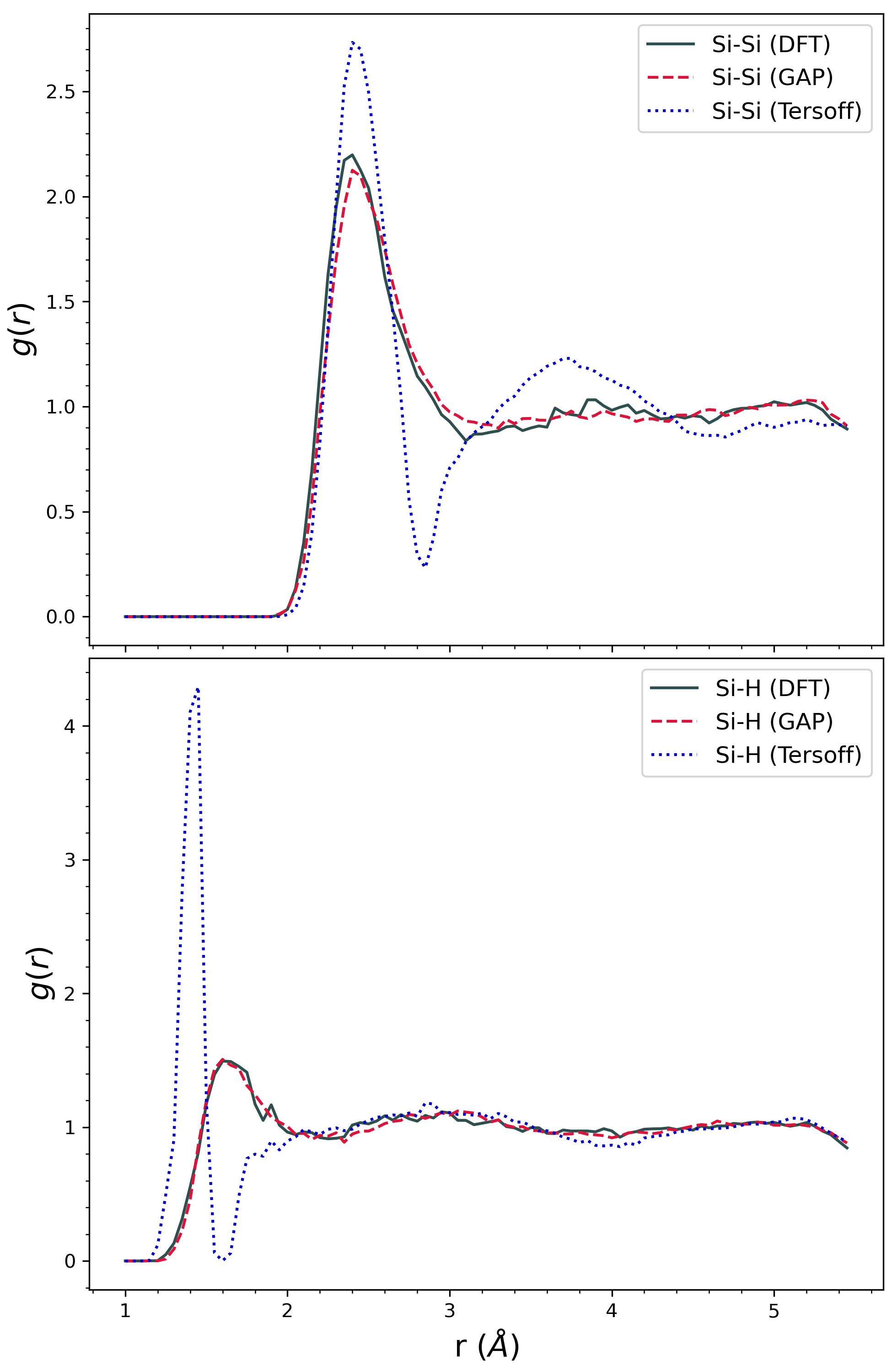}
\caption{Partial pair correlation functions of equilibrated liquid Si infused with H, with a density of 2.58 g/cm$^3$, and $T$=2000K. Top: Si-Si; bottom: Si-H. Comparison provided between the GAP, a Si-H Tersoff potential, and DFT.}
\label{fig:liquidRDF}
\end{figure}

To simulate the structure of liquid Si infused with H, we performed constant number/volume/temperature (NVT ensemble) molecular dynamics simulations as implemented in the LAMMPS software package, built with QUIP package support \cite{LAMMPS, LAMMPSLINK, QUIP}. Separate simulations were carried out using the Si-H GAP and a Si-H Tersoff potential \cite{Tersoff1989, deBritoMota} to provide a comparison basis. Each simulation started with a cubic supercell of side-length 5.26 $\mathrm{\AA}$, containing 64 Si atoms and 8 H atoms placed at random, corresponding to a density of 2.58 g/cm$^3$ and to an approximate H at. concentration of 12\%. The locations of the atoms were optimized with the interatomic potential, before equilibrating at T=2000K for 150,000 timesteps of 0.5 fs duration. Structural data was gathered over an additional 10,000 timesteps of 0.25 fs. 

The same observables were also determined by DFT Born-Oppenheimer molecular dynamics (DFT-BOMD) simulations using Quantum Espresso, or gathered from the literature where available. Just as above, the starting configuration consisted of 64 Si and 8 H atoms located at random non-overlapping positions, in a cubic supercell of side-length 5.26 $\mathrm{\AA}$. Simulations consisted of equilibrating the structure at T=2,000K over 100,000 timesteps of 0.25 ps, using the Verlet algorithm, rescaling the velocities at every step to keep the temperature fixed at T=2,000 K. After equilibration, structural data was gathered over an additional 6,000 time steps of 0.25 ps.

Fig. \ref{fig:liquidRDF} shows the Si-Si and Si-H partial pair correlation functions $g(r)$ computed by these three methods, sometimes also referred to as radial distribution functions. Visibly, the partial pair correlation functions $g(r)$ computed with our Si-H GAP are in excellent agreement with the results of the DFT computation: both the peak locations and peak heights exhibit high quality matching. In contrast, the partial pair correlation functions $g(r)$ determined with the Tersoff potential deliver much poorer correspondence with DFT.

\begin{figure}[t]
\includegraphics[width=0.95\columnwidth]{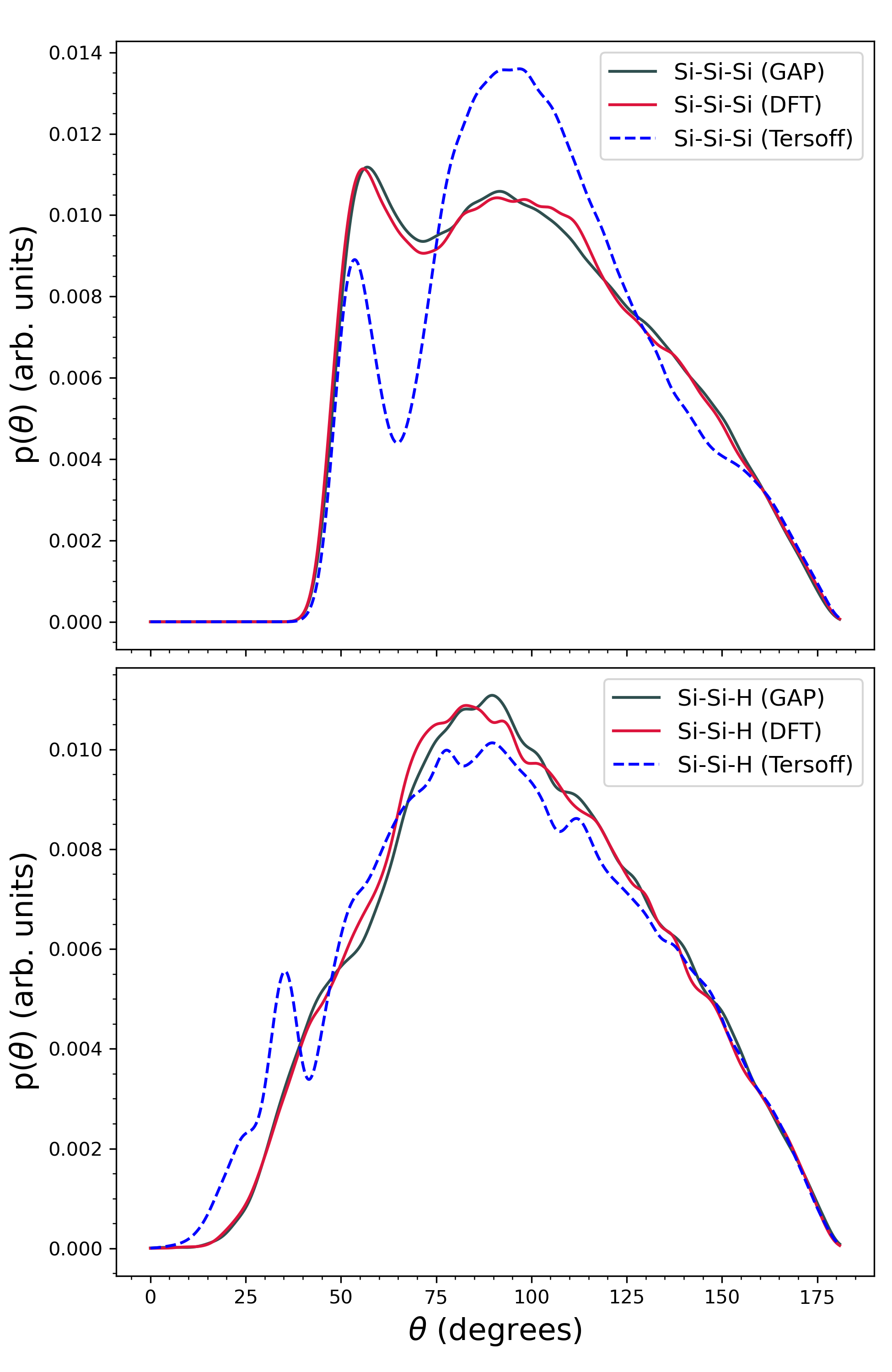}
\caption{Angular distribution functions, or $g_3(r,\theta)$, of equilibrated liquid Si infused with H, with a density of 2.58 g/cm$^3$, and $T$=2000K. Top: Si-Si-Si; bottom: Si-Si-H. Comparison provided between the Si-H GAP, a Si-H Tersoff potential, and DFT.}
\label{fig:liquidADF}
\end{figure}

Fig. \ref{fig:liquidADF} shows the bond-angle distribution functions $p(\theta)=g_3(r_\mathrm{cutoff},\theta)$, also known as the angular distribution functions. The cutoff bond length $r_\mathrm{cutoff}$ for $g_3(r_\mathrm{cutoff},\theta)$ is taken to be the radius corresponding to the first minimum of the partial pair correlation functions $g(r)$ beyond their initial peaks, 3.1 $\mathrm{\AA}$ for Si-Si, and 2.2 $\mathrm{\AA}$ for Si-H. Just like with the partial pair correlation functions $g(r)$, the computational results with our Si-H GAP for both the Si-Si-Si bond angle distribution and the Si-Si-H bond angle distribution are in excellent agreement with the DFT computations. They reproduce all notable features of the DFT results with high quality. As before, the calculations using the Tersoff potential track the DFT results with a substantially inferior quality.

\begin{figure}[t]
\includegraphics[width=0.7\columnwidth]{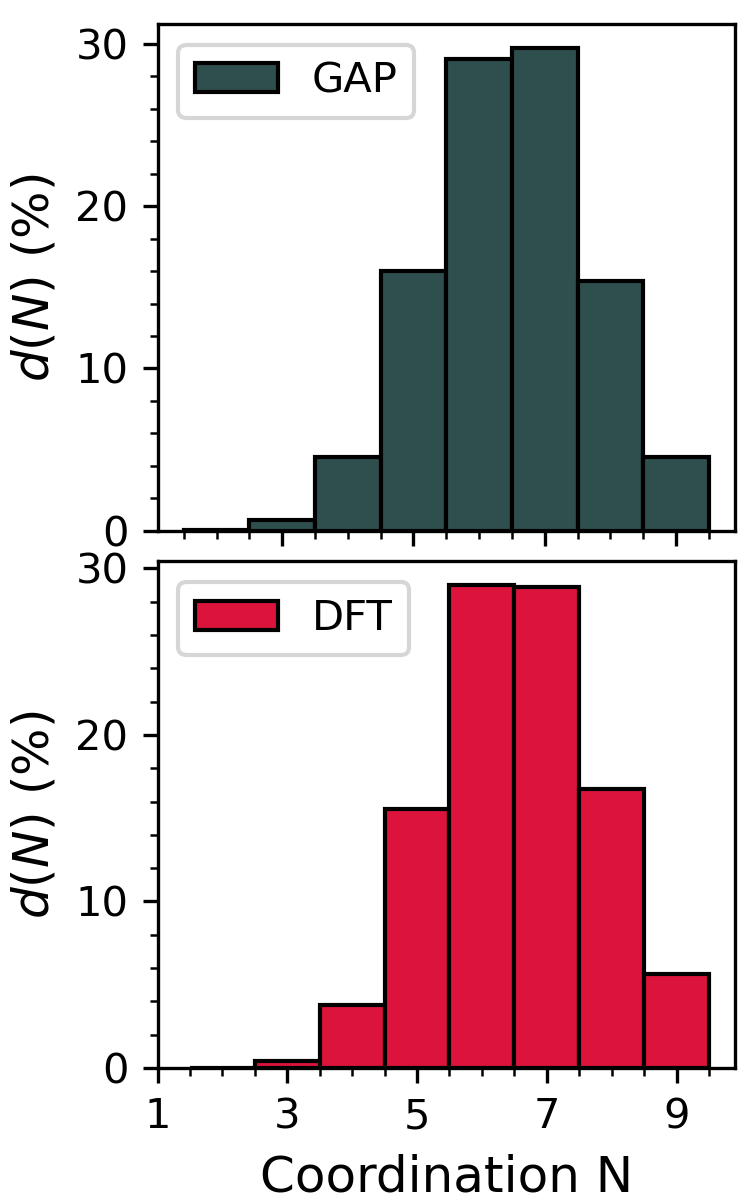}
\caption{Averaged coordination statistics of Si atoms in equilibrated liquid Si infused with H, with a density of 2.58 g/cm$^3$, and $T$=2,000K. Top: GAP; bottom: DFT.}
\label{fig:liquidCoordination}
\end{figure}

Finally, Fig. \ref{fig:liquidCoordination} shows the coordination statistics of the Si atoms. Results are only shown for the Si-H GAP and DFT. Here the coordination includes the neighboring Si and H atoms both. The coordination shell for each atomic species is again defined using the first minimum of the corresponding partial pair correlation function. On this front the Si-H GAP once again delivers excellent agreement with the DFT. To place these observables in context, we refer to Ref. \citenum{carParrinelloLiqSi} that provides these same observables for pure liquid Si.

\subsubsection{Amorphous phase}

\begin{figure}[b]
\includegraphics[width=0.95\columnwidth]{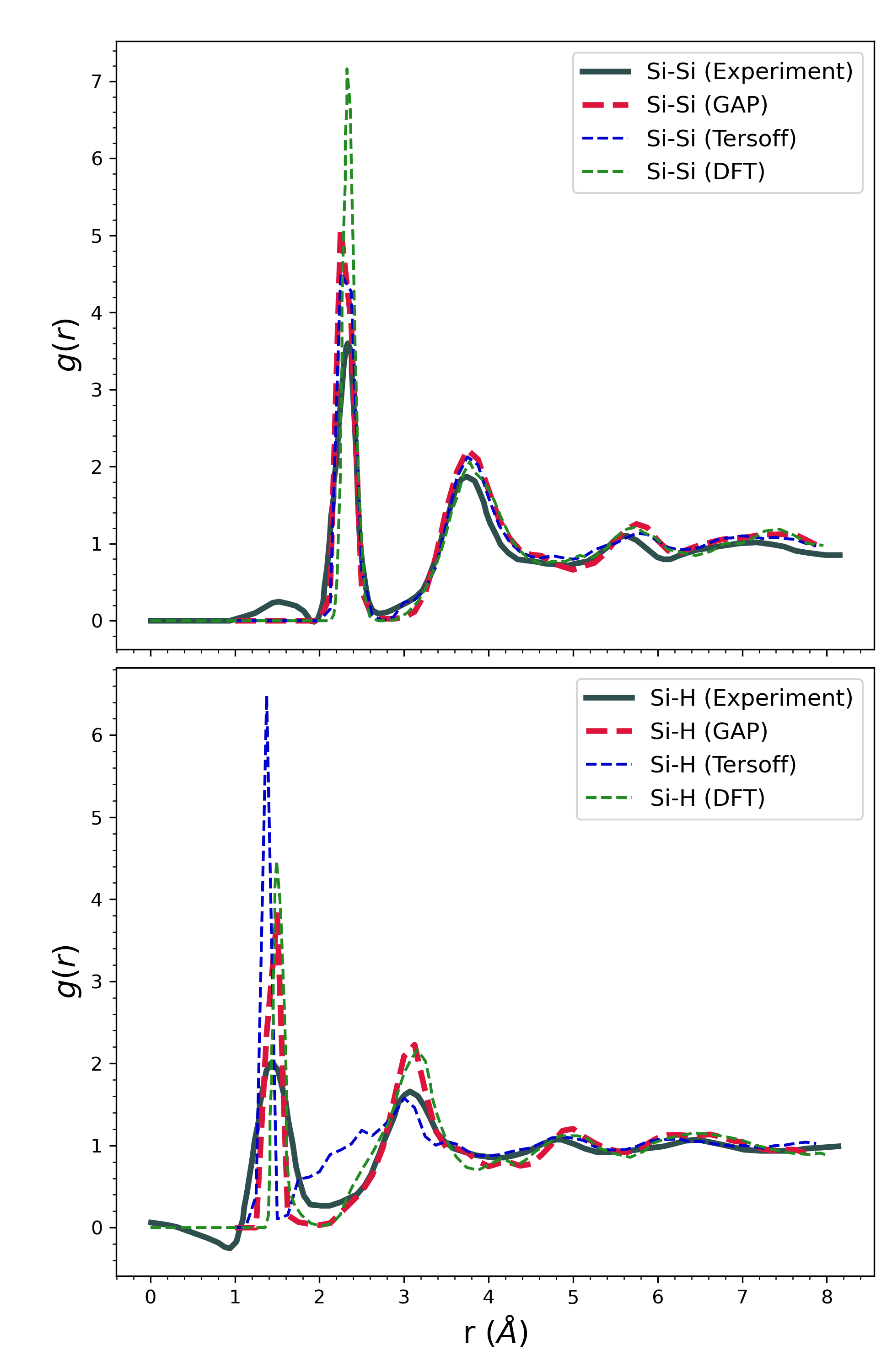}
\caption{Partial pair correlation functions of a-Si:H. Top: Si-Si; bottom: Si-H. Results provided for GAP and Tersoff simulations, and reference results generated by a neutron scattering experiment (reproduced as published) \cite{bellisent} and DFT calculations \cite{cooling3} are shown for comparison.
}
\label{fig:amorphousRDF}
\end{figure}

Next, we simulated the structure of a-Si:H. To this end, we performed both constant volume (NVT ensemble) and constant pressure (NPT) molecular dynamics simulations, again as implemented in the LAMMPS software package with QUIP package support. As before, we performed the calculation using our Si-H GAP, and then performed the same calculation using a Si-H Tersoff potential \cite{Tersoff1989, deBritoMota} to provide a point of comparison. Each simulation started with the same general procedure as the previous, liquid phase section, except with a larger supercell containing 216 Si atoms and 28 H atoms, and an initial density of 2.3 g/cm$^3$. Once the liquid structures were equilibrated, the same general procedure was followed for each potential, except the Tersoff calculations used cooling rates which were reduced by a factor of two for further accuracy.

\begin{table*}[t]
\begin{center}
 \begin{tabular}{||m{4.5cm} | m{1.8cm} | m{2.0cm} | m{2.0cm} | m{1.5cm} | m{1.5cm} ||} 
 \hline
 a-Si:H structure (H at. \%) & $N_c$ & $r (\mathrm{\AA})$ & $\sigma_r (\mathrm{\AA})$ & $\theta_{\mathrm{Si-Si-Si}}$ & $\sigma_{\theta, \mathrm{Si-Si-Si}}$\\ [0.5ex] 
 \hline\hline
 Si-H GAP (11.5\%) & 3.98 $\pm$ 0.01 & 2.376 & 0.047 & 109.2 & 10.7\\ 
 \hline
 Tersoff (11.5\%) & 4.14 $\pm$ 0.01 & 2.392 & 0.077 & 109.0 & 13.2\\ 
 \hline
 Jarolimek (DFT, 11.1\%) \cite{cooling3} & 3.89 $\pm$ 0.03 & 2.377 & 0.049 & 108.9 & 13.6\\
 \hline
 Filipponi (PECVD, 8\%) \cite{filipponi1989} & 3.88 $\pm$ 0.12 & 2.35 $\pm$ 0.01 &  &  &  \\ 
 \hline
 Vignoli (PECVD, 12\%) \cite{vignoli2005} & 3.71 $\pm$ 0.07 & 2.37 $\pm$ 0.04 &  &  & 8.7\\ 
 \hline
 Wakagi (sputtering, 12.8\%) \cite{wakagi1994} & 4.0 & 2.363 $\pm$ 0.004 & 0.038 $\pm$ 0.008 &  & 9.3\\ 
 \hline
 Kail (various, 12-15\%) \cite{kail2010} &  &  &  &  & 9.0-9.7\\ 
 \hline
\end{tabular}
\caption{Short-range order of the a-Si:H produced by the Si-H GAP compared to Tersoff, DFT, and PECVD and sputtering (experiment) methods. $N_c$ is the average coordination of the Si atoms. $r$ and $\sigma_r$ are the mean and the deviation of the Si-Si bond length. $\theta_{\mathrm{Si-Si-Si}}$ and $\sigma_{\theta, \mathrm{Si-Si-Si}}$ are the mean and the deviation of the Si-Si-Si bond angle distribution. The coordination and means are gathered at $T=300$ K, while the deviations were measured on a snapshot of an optimized structure. }\label{tab:fourth}
\end{center}
\end{table*}

The details of the GAP simulation procedure were as follows. We formed a liquid Si:H phase structure  at T=2,000K. Once the liquid structure was equilibrated, it was cooled in the NVT ensemble down from T=2,000K to T=1,500K at a rate of $10^{13}$ K/s with a timestep of 1 fs. Further equilibration was then performed at T=1,500K for 100 ps, before the structure was cooled down to T=500K at a rate of $10^{12}$ K/s. Both of these steps were also performed in the NVT ensemble. To collect structural data, the structure was relaxed to the local energy minimum in atomic positions, then equilibrated over 20,000 timesteps of 1 fs in the NPT ensemble at 0 pressure and T=500K, followed by gathering the structural data over an additional 10,000 timesteps. This procedure resulted in a-Si:H with a mass density of approximately 2.22 g/cm$^3$, consistent with PECVD films containing 12 at. \% H \cite{smets2003}. 

For this comparative analysis, the DFT-BOMD data was gathered from Ref. \cite{cooling3}. This paper performed a very extensive and therefore computationally expensive study, thereby producing very high quality data. Finally, the comparison base was extended by including experimental data, acquired by neutron scattering measurements \cite{bellisent}.

The partial pair coordination functions $g(r)$ are presented in Fig. \ref{fig:amorphousRDF}. Results are given for the Si-Si and the Si-H partial pair correlation functions $g(r)$. The key features of the Si-Si $g(r)$ are the three strong peaks at about 2.4 $\mathrm{\AA}$, 3.8 $\mathrm{\AA}$, and 5.6 $\mathrm{\AA}$, corresponding the first, second and third neighbor peaks. Similarly, the key features of the Si-H $g(r)$ are the four peaks at about 1.5 $\mathrm{\AA}$, 3.1 $\mathrm{\AA}$, 4.9 $\mathrm{\AA}$, and 6.5 $\mathrm{\AA}$, corresponding to preferential Si-H separations. Remarkably, the Si-Si and the Si-H partial pair correlation functions, produced with our Si-H GAP, both achieved excellent agreement with the reference $g(r)$ of the DFT calculations once again, matching peak locations and heights. In fact, the Si-H GAP correlation functions achieved slightly better agreement with the experiments than the DFT correlation functions. The Si-Si partial pair correlation function of the Tersoff potential notably improved compared to the liquid-Si:H case, but the Si-H partial pair correlation function continued to provide only a poor fit to the DFT and experimental results.

We extended and strengthened our validation of the Si-H GAP by performing further structural measurements. Table \ref{tab:fourth} shows time-averaged measurements of the following quantities. The average coordination of the Si atoms $N_c$, the mean Si-Si bond length $r$ and its standard deviation $\sigma_r$, and the mean angle $\theta_{Si-Si-Si}$ of the Si-Si-Si bond angle distribution, and its standard deviation $\sigma_{\theta,Si-Si-Si}$. (The Si-Si-Si bond angle distribution is Gaussian in the case of a-Si:H, so its standard deviation is a well defined quantity). When taking these measurements, the cutoff bond lengths were taken to be the first minimum values of the partial pair correlation functions in Fig. \ref{fig:amorphousRDF} after their initial peaks, 2.75 $\mathrm{\AA}$ for Si-Si, and 1.9 $\mathrm{\AA}$ for Si-H. 

As before, the validation and utility of the Si-H GAP is demonstrated by a comparison to corresponding measurements with three relevant techniques: MD using Tersoff potentials, DFT, and experimental measurements, of the PECVD and sputtering type. The atomic concentrations of H in the referenced a-Si:H measurements varied from 8 to 12.8\%. In all cases the coordination and means were calculated on atomic configurations held at $T = 300$K. Also shown in Table \ref{tab:fourth} are measurements of the static disorder, namely the deviation of the Si-Si bond lengths and Si-Si-Si bond angles, measured on a snapshot of an optimized structure. 

The key message of Table \ref{tab:fourth} is that MD calculations with Si-H GAP reached the precision of DFT, and thus produced results consistent with the experiments for 4 of the 5 observables, whereas Tersoff-based MD showed notable variance relative to DFT and experiments. Intriguingly, for the fifth observable, the width of the bond angle distribution $\sigma_{\theta,Si-Si-Si}$ the DFT results deviated from the experiments considerably, and in this case the Si-H GAP result reproduced the experiments even better than DFT. we also measured the Si-Si-H bond-angle distribution, where the mean bond angle was found to be $105.9 \pm 6.0\degree$, and the average Si-H bond length was found to be $1.53\pm0.03 \,\mathrm{\AA}$.

We also note that the coordination of the Si atoms $N_c$ was found to be very close to 4, which indicated that very few dangling bonds were present in the GAP supercell even at $T=300$ K. This is compelling evidence that even this limited concentration of H of 12.5\% was able to passivate Si dangling bonds so efficiently that only 2\% of all bonds remained unsatisfied. 

\begin{figure}[t]
\includegraphics[width=0.95\columnwidth]{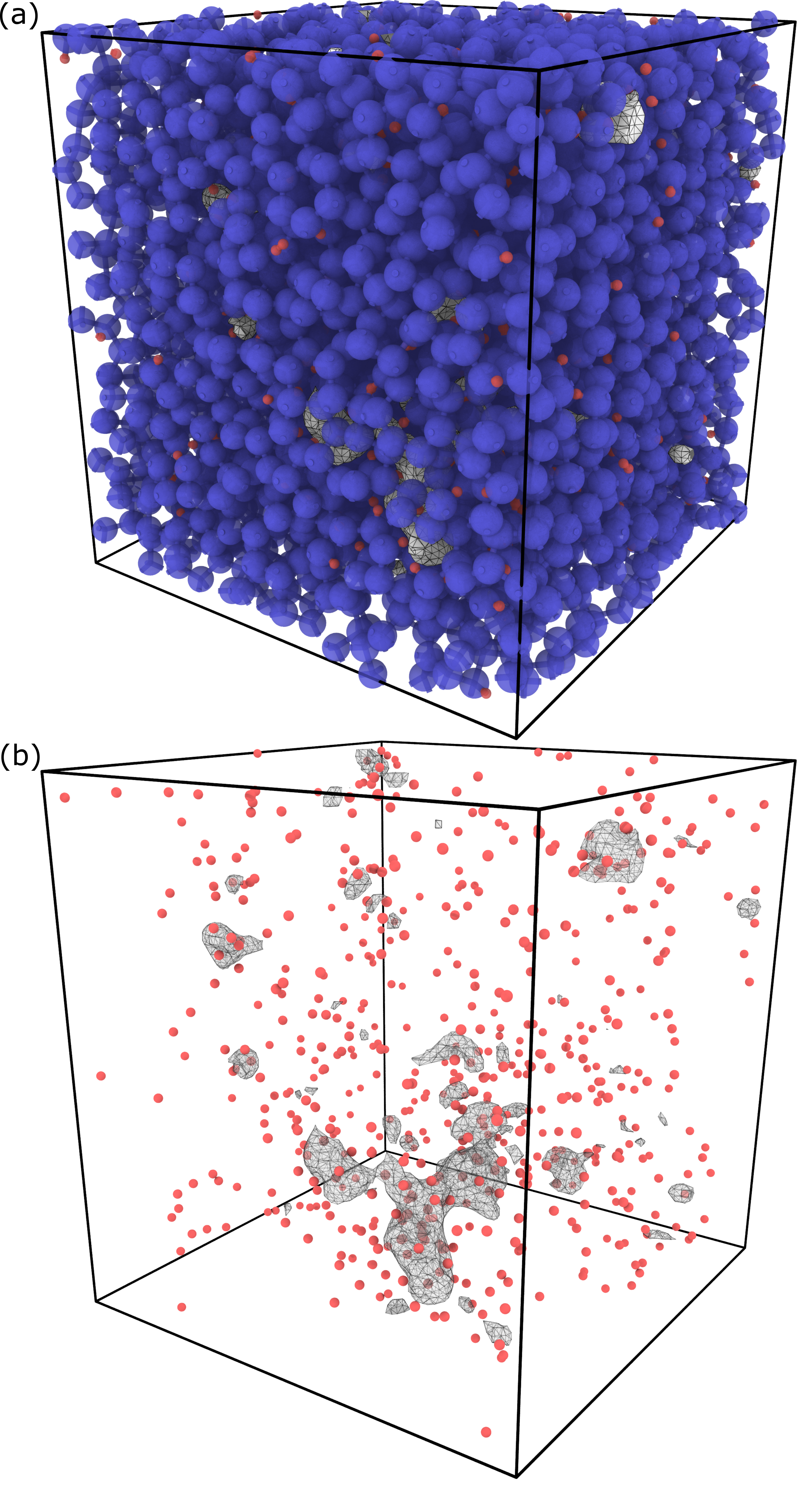}
\caption{(a) Ball-and-stick rendering of a structural model of a-Si:H, containing 4096 Si atoms and 558 H atoms. Si atoms are presented in blue, and H atoms are presented in red. The larger voids which are present in the atomic configuration are highlighted here with white surface meshes. (b) A visualization of the distribution of H atoms and voids within the supercell of the structural model. The voids here are represented with transparent white meshes.}
\label{fig:voidVisual}
\end{figure}

Next, we calculated the lowest achievable ``excess energy'' of the Si:H supercell. This excess energy was defined as the energy difference of the supercell compared to a same-size supercell containing hydrogenated \textit{crystalline} silicon with the same number of Si and H atoms, where the H atoms were placed at the tetrahedral interstitial sites. 
When both energies were computed with Si-H GAP alone, the excess energy was 0.20 eV/atom. When both structures were additionally relaxed with DFT, the excess energy was 0.09 eV/atom. This latter excess energy is well within the measured a-Si:H experimental range of 0.06 to 0.13 eV/atom \cite{excessenergy3}. As expected, it is lower than the excess energies of 0.13 to 0.14 eV/atom of pure amorphous silicon structures that we generated using the Si-only GAP \cite{PRX} for a previous work \cite{soldeg}, as the presence of the H atoms relieved some of the atomic strain. 

As a final demonstration of the utility of our ML-based Si-H interatomic potential GAP, we conducted a study of extended defects, as those play an important role in the description of a-Si. Using the same procedure as above, we created an  large a-Si:H supercell that contained 4096 Si atoms and 558 H atoms. This system size is prohibitively expensive for DFT-BOMD, since the computational resources required by DFT-BOMD scale with $N^3$, while this size is comfortably attainable for MD utilizing the Si-H GAP. We created this large supercell using the same melt-quench procedure as described for the smaller supercells. A ball-and-stick rendering of the large supercell is presented in Fig. \ref{fig:voidVisual}a. We checked that the partial pair correlation functions and other short-range order statistics are almost identical for Si$_{216}$H$_{28}$ and for Si$_{4096}$H$_{558}$ supercells, so we do not separately show them here. Instead, we highlight features that manifest on medium- to long-range scales, and thus can \textit{only} be studied in such large supercells: the void structure of a-Si:H. 

We characterized the void structure. First, we used the Zeo++ package \cite{zeo1} to sweep a set of Monte-Carlo sampled points in our structural model and constructed the largest spheres which could be centered on each point without touching any atoms \cite{zeo2}. We then switched to the Ovito visualization package to analyze whether these individual spheres overlapped/joined with other spheres. When this happened, we constructed the union of these spheres as a definition of larger voids. An example is visible at the lower-front of Fig. \ref{fig:voidVisual}a and Fig. \ref{fig:voidVisual}b. Fig. \ref{fig:voidVisual}b offers an alternative visualization, where we only show the hydrogen atoms (in red), and the larger voids, in grey.

\begin{figure}[t]
\includegraphics[width=0.95\columnwidth]{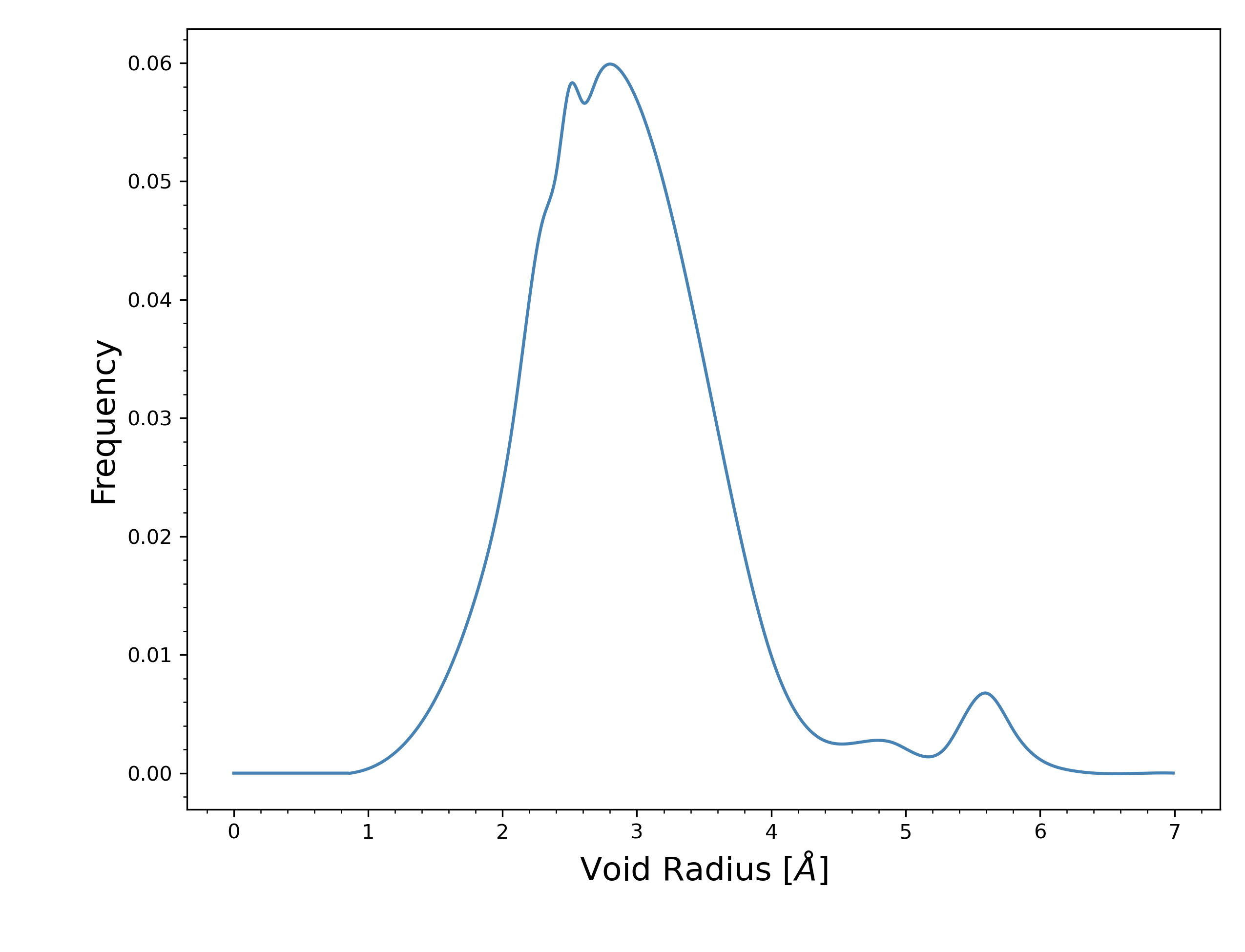}
\caption{The void radius histogram of the a-Si:H structural model shown in Fig. \ref{fig:voidVisual}, as calculated by Zeo++.}
\label{fig:voidHistogram}
\end{figure}

Fig. \ref{fig:voidHistogram} shows the void radius histogram, or pore-size distribution function, for the structural model of Fig. \ref{fig:voidVisual}. Conspicuously, the majority of the voids present in the structure have radii between 2 and 4 $\mathrm{\AA}$, which is consistent with the void structure observed in WWW-generated a-Si:H configurations \cite{strubbeVoids}. This range of void radii approximately corresponds to the size of mono- or di-vacancies in c-Si. There are also a few larger voids present in the structure with radii between 5 and 6 $\mathrm{\AA}$. Fig. \ref{fig:voidVisual}b shows that these larger-radius voids can and do have irregular shapes, often referred to as ``nanovoids'' \cite{smets2003}. In the studied H concentration range around 12\%, hydrogen released the structural strains very efficiently, and this relaxation caused most of the defects to shrink into mono- and di-vacancies, with only a few nanovoids left. Experiments studying the nature of voids in a-Si:H have found that nanovoids only begin to dominate above 14 $\pm$ 2 at. \% H \cite{smets2003}. 

\section{Conclusion}

In this paper, we reported on the delvepment of a silicon-hydrogen Gaussian Approximation Potential with a SOAP kernel,  Si-H GAP. We trained this potential over 27 iterative rounds, simulating a wide range of phases, structures, and thermal protocols, while fitting energies, forces and stresses to the results of DFT calculations in each round. The training set was expanded in each round only as required by an adaptive protocol. This ``adaptive training'' enables the economic use of  computational resources.

The Si-H GAP was not only able to closely match DFT measurements of microscopic quantities such as the energies, forces and virial stresses, but it was also able to reproduce structural characteristics including partial pair correlation functions, bond angle distributions, bond length distributions, and average coordination numbers. 

For context and comparison, we also performed MD simulations based on the Tersoff potential and calculated the same quantities. The comparison of the parametric Tersoff and non-parametric ML-based Si-H GAP results demonstrated the improvement that is obtained by adopting the non-parameteric ML model. While the impressive increase of accuracy of the Si-H GAP comes at a cost of computational complexity and speed over traditional parameteric interatomic potentials, Si-H GAP achieves DFT-level accuracy with while still being many orders of magnitude faster than DFT.

The Si-H GAP developed here is limited in a few ways. (i) It was not developed to be a general-purpose interatomic potential, and thus did not include structural phases beyond the key phases which are needed in order to reliably produce experimentally accurate amorphous Si:H. For instance, it did not include all of the various surface reconstructions of diamond Si. Fortunately, future training could simply add additional structures to the training set, should this be desired, and thus the potential can be adapted as needed in future work. (ii) Long-range interactions beyond 5 $\mathrm{\AA}$ for Si and 3.5 $\mathrm{\AA}$ for H were not included. This limits the maximum accuracy of the potential, as it runs into locality limits. For a discussion of this phenomenon, see Ref. \citenum{C_GAP}. Properly including a long-range interactions and integrating them with the short-range interactions is still an outstanding problem within the GAP framework, although simple dispersion and Coulomb interactions are easy to add on, as has been done before in Refs. \citenum{GAP1} and \citenum{Phosphorus_NatComm2020,Rowe_CGAP2020}.
(iii) The training database was assembled by hand, and did not include any automated procedures for constructing the database. Using an active learning approach to create and manage the training set would reduce the cost of adaptively expanding the training set, as DFT would only be performed on an ``as needed'' basis, and would speed up the training of the potential.

Those issues aside, the merits of the presented potential are clear.  Since in electro-optical applications, and in most experimental studies amorphous silicon structures typically contain some amount of hydrogen, our Si-H GAP can prove to be very useful to enable simulation studies with unprecedented accuracy and utility.

\section*{Acknowledgements}

We acknowledge useful and inspiring discussions with Chase Hansen, Mariana Bertoni, and Salman Manzoor. This research was supported by DOE SETO grant DE-EE0008979.

\section*{Software Availability}

The GAP suite of programs is freely available for non-commercial
use from {\tt http://github.com/libatoms/GAP}. The Quantum Espresso software
package is freely available from {\tt www.quantum-espresso.org}.
The LAMMPS software package is freely available from
{\tt lammps.sandia.gov}. Zeo++ is freely available from {\tt www.zeoplusplus.org}. 

%

\end{document}